# WEB SERVICES SUPPLY CHAINS: A LITERATURE REVIEW

Krithika V[1], Dr. Arshinder Kaur[2] and Dr. K. Chandra Sekaran[3]

[1]Department of Management Studies, Indian Institute of Technology Madras, India

krithika1288@gmail.com

[2] Assistant professor, Department of Management Studies, Indian Institute of Technology Madras, India

arshinder@gmail.com

[3]Professor, Department of Computer Science Engineering, National Institute of Technology, Surathkal, India

kchnitk@gmail.com

Abstract:

*The aim of this review paper is to bring into light a potential area i.e., web services supply chains for research by analyzing the existing state of art in this. It is observed from the review process that there seems to be much less work done in the area of web service supply chains as compared to e-commerce and product oriented service supply chains. The service quality assurance models, end to end Quality of Service (QoS) models, attempts made to QoS attributes are also found to be from individual perspectives of participating entities in a service process rather than a collective perspective considering individual QoS attributes rather than multiple QoS attributes. In light of these gaps we highlight the comparison between product oriented and pure online/ web service supply chains, a need for quality driven optimization in the web services supply chains, perceived complexities in the existing work and propose a conceptual model.*

I.     Introduction

The globalization era has bought almost all services to the internet environment with multiple providers providing the same service in different ways. The same functionality is being offered by multiple providers so it becomes necessary to differentiate one self from the crowd in order to thrive against the cut throat competition from the service provider's view. On the other hand it is absolutely essential for the client to select a service provider who meets not only the functional requirements of the client but also provides the best possible quality of service (QoS) to the customers. The services industry has contributed approximately 55.3% to the GDP of India in the year 2009. The service sector employs about 34% of the labor force in India [CIA report]. The exponentially growing service industry has started making its global presence through the internet enabled online environment. The web environment has made almost all kind of services available online be it a simple movie ticket booking or the most complex processes such as outsourcing, funds transfer and others. The online services are special kind of services as compared to the static or product supply chain which is more offline or the semi online service supply chains in which part of the service is offline and partly online.

The objective of this paper is to present a literature review about web services supply chains and the non – functional or QoS attributes consideration in them. The web service non – functional attributes considered in the literature so far, the methodologies used the perspectives of consideration and the topics considered till now. The outcome is to point out the research gap as far as web services supply chains and the QoS aspects of web services supply chains are considered and possible future research work.

II.     Online service supply chains and the web service ecosystem

A. Defining service supply chains:

The service supply chain literature till now considers static services/services allied to a manufacturing product and semi – online services where a part of service takes place online and remaining takes place offline. The pure online services / web services on the other hand involve service delivery by the dynamic collaboration of multiple entities in the internet environment. The simple definition of web services come from [1] which defines web services as a service offered as a software. The collaboration is many times not a pre- defined or fixed one it involves selection from a pool of service providers and intermediate providers based on a virtual run time contract. The service selection, service providing and acceptance are instantaneous and dynamic. The service supply chain is defined as, an integration of a series of entities (individual person, organization, enterprise) to provide personalized service directly or indirectly [2]. This definition when coupled with the internet environment makes the service supply chain more dynamic. The dynamism makes these supply chains more different from the service supply chains definitions in the literature. The advent of advanced concepts like service oriented architecture SOA and web services have made the dynamic collaboration possible [3].The indulgence of SOA and web services resulted in the web service ecosystem defining the participants of online service supply chains [4]. The web eco system has three main entities namely the service provider, intermediate and the client/ customer.

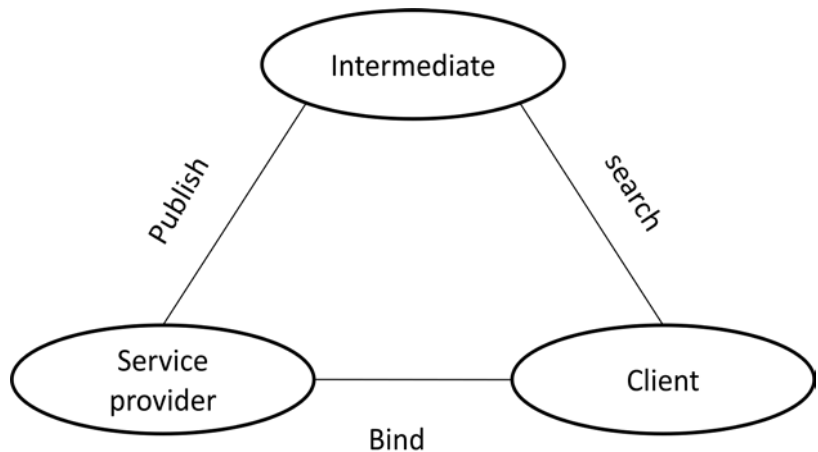

Fig 1: Web service ecosystem

The intermediate can be either other service providers or just brokers. The brokers can be of two types namely forwarding broker and match making broker. The forwarding brokers play the role of plumbing between the service provider and the client through which the client communicates with the service provider to get the service. The match making broker on the other hand just act like match makers in the real world they just connect the service providers and clients acting like a common connecting point [5]. The performance of the eco system depends on all the three entities but in case of matchmaking broker type ecosystem the performance of the broker will not affect the ecosystem much because the broker just acts as the connecting point.

B. Comparison of service supply chains with manufacturing supply chains

The properties that differentiate service supply chains from product oriented / manufacturing supply chains are,

i. Perish ability: A service perishes once it is not consumed on time i.e the service cannot be reused later e.g. A flight's empty seat after the take off of the airplane is a loss since it has already perished

ii. Simultaneity: Services must be consumed simultaneously at the time of production since they cannot be inventoried or stored

iii. Intangibility: Services are mostly not physical i.e. tangible they are intangible because they are either experience oriented or meant for individual demands.

iv. Non – transferability: The services are non transferable due to the lack of ability to be inventoried.

These characteristics are based on the literature [2] on service supply chains till now which is very limited in nature.

III. Research methodology

A. Journals and Conferences

The literature review presented here is based on the exploratory study of literature from various prestigious international journals and conferences. The journals and conferences considered for the literature review include the following,

| Publication | Total | Conference Papers | Journal Papers | % over total number of papers |
|---|---|---|---|---|
| IEEE | 55 | 44 | 11 | 44.35 |
| Elsevier | 8 | 0 | 8 | 6.45 |
| Emerald | 4 | 0 | 4 | 3.23 |
| International Journal of web service research | 5 | 0 | 5 | 4.03 |
| International Journal of web service practices | 4 | 0 | 4 | 3.23 |
| Journal of Computers and System Sciences | 4 | 0 | 4 | 3.23 |
| Journal of Simulation | 3 | 0 | 3 | 2.42 |
| Information Technology Journal | 5 | 0 | 5 | 4.03 |
| Other journals / Conferences | 36 | 4 | 32 | 29.03 |
| Total | 124 | 48 | 76 | |

Table1. List of references

The contribution in terms of the number of journals and conferences include areas such as web service's non-functional aspects, method of consideration of non- functional attributes, web services and service oriented architectures in enabling service supply chains, service oriented supply chains.

B. Classification topics:

The topics are considered in such a way so as to investigate the relationship among the enablers of service supply chains, the non-functional aspects of online services considered so

| Topics | Total | Conference papers | Journal Papers | % over total number of papers |
|---|---|---|---|---|
| Broker based web services selection with QoS | 4 | 3 | 1 | 5.71 |
| Business service networks (SOA and web services for supply chains) | 5 | 3 | 2 | 7.14 |
| Web service selection with QoS | 9 | 3 | 6 | 12.86 |
| Web service discovery with QoS | 3 | 1 | 2 | 4.29 |

far in the literature thus moving from lower to higher level of abstraction while classifying the literature. The classification we present concentrates more on the web services and their non- functional properties/ QoS aspects and service supply chains.

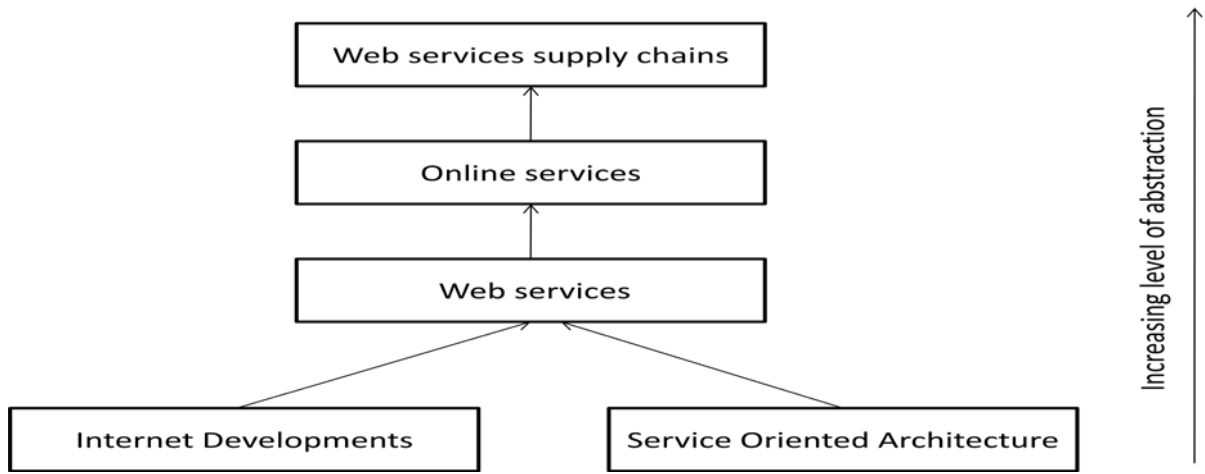

Fig 2. Classification based on topic

Table 2. Classification based on topic

| Web service QoS monitoring/support | 5 | 5 | 0 | 7.14 |
| --- | --- | --- | --- | --- |
| Web service composition with QoS | 14 | 4 | 10 | 20.00 |
| Non - functional properties / QoS of web services | 13 | 12 | 1 | 18.57 |
| Survey/ Classification of QoS properties | 3 | 3 | 0 | 4.29 |
| Service oriented architecture (SOA) for supply chains | 9 | 5 | 4 | 12.86 |
| Service supply chains | 3 | 3 | 0 | 4.29 |
| Supply chains as collborative networks | 2 | 2 | 0 | 2.86 |
| Total | 70 | 44 | 26 | |

The literature review process how ever considers the literature related to service oriented architecture's (SOA) role in enabling web services [6], [7], [8], [9], [4] SOA's applicability to supply chains in enabling automation, collaboration and integration of supply chains [10], [11], [12], [13], [14], and so the classification is done on topics in each aspect as shown in fig 2.

The literature classification based on various topics considered so far shows that the research has concentrated on web service composition and selection while considering QoS predominantly. The non functional properties or QoS aspects of web services have gained importance because it is necessary to differentiate a web service which is in turn a pure service [1] based on quality when multiple providers of same functionality are present. The automation of manufacturing supply chain has gained importance due to the expansion of industries and need to cater to global demand and compete globally. The advent of SOA and web services has resulted in middle wares and broker based architectures to be routes of QoS negotiation between service providers and clients. There is scanty amount of work in the area of service supply chains in general irrespective of semi- static or pure online services supply chains and their collaboration. The combination of SOA and web services has started to pave way for a new class of research study called the business service networks that majorly collaborate in the internet environment. How ever the research in this area is very less as it can be seen from Table 2.

C. Classification based on Perspectives

The use of SOA and web services brings into light the need to consider the service supply chain entities namely the service provider, intermediaries and the client [4]. The perspectives all these three participating entities must be considered and the benefit for each of these entities must be taken into account. The perspective of each entity must be analyzed and maximum gain for each must be identified and balanced without affecting the interest of another entity. This makes it necessary to look at the literature interns of the perspectives of the participating entities to get a clear picture of the state of the art research in this dimension.

| Perspective | Total | Conference Papers | Journal Papers | % over total number of papers |
|---|---|---|---|---|
| Client / Customer view | 28 | 12 | 16 | 35.44 |
| Service provider view | 34 | 24 | 10 | 43.04 |
| Middle wares | 5 | 5 | 0 | 6.33 |
| Mutual view (service provider and client) | 3 | 2 | 1 | 3.80 |
| Generic (applicability of concepts) | 9 | 6 | 3 | 11.39 |
| Total | 79 | 49 | 30 | |

Table 3. Classification based on perspective

The classification clearly shows that the research till now pertains to one of the participating entities predominantly the clients and the service providers. The research works considering the participating entities' effect on each other, the conflict of interests that might arise, how the performance of one entity affects the collective performance, the optimal gain possible for all the entities and other related questions are scarce. The middle wares how ever have started gaining research focus but the mutual consideration is still taking a back seat in terns of the research work.

D. Classification based on methods

The method used in a research decides the perspective and objective of a research work. The methods range from very old ones to brand new ones with many new dimensions to the research problem and tradeoffs for each method used. The classification based on methods is shown below in table 4. The classification clearly shows that conceptual work, ontology and frame works have got lot of research attention. There is how ever equal amount of mathematical formulation and prototyping/ implementation methods used to prove the concepts proposed. The use of automata theory to handle time properties is gaining importance. The extension of this is the latest use of petrinets to verify time properties of online services. The need to incorporate client requirements and suggestions has started getting attention with the use of methods like quality function deployment, fuzzy logic and neural networks approaches that help to convert subjective user inputs to objective values that can be used for further analysis to take appropriate actions.

| Method | Total | Conference Papers | Journal Papers | % over total number of papers |
|---|---|---|---|---|
| Petri nets | 4 | 1 | 3 | 5.33 |
| Fuzzy logic/ Neural networks | 4 | 2 | 2 | 5.33 |
| Quality Function Deployment | 2 | 1 | 1 | 2.67 |
| Timed Automata | 3 | 3 | 0 | 4.00 |
| Ontology | 13 | 9 | 4 | 17.33 |
| Architectures and Frame work | 20 | 13 | 7 | 26.67 |
| Graphs | 2 | 1 | 1 | 2.67 |
| Conceptual | 10 | 6 | 4 | 13.33 |
| Mathematical formulation and analysis | 8 | 5 | 3 | 10.67 |
| Implementation and Prototyping | 9 | 5 | 4 | 12.00 |
| Total | 75 | 46 | 29 | |

Table 4a. Classification based on methods

The research how ever points out that the collaborating service entities in the web environment form a logical graph rather than a traditional client server link [15] but the research methods with graphs is very scanty compared to the other methods that have been used till now. The combinations of some methods such as fuzzy logic and quality function deployment [16] are used together to get better results. The advent of programming languages and ease of programming has led to the increase in prototyping and implementation of concepts easily. The operations research (OR) methods applied in this research area are shown below in table 4b.

| Operations research methods used | Total | Conference Papers | Journal papers | % over total number of papers |
|---|---|---|---|---|

| Genetic Algorithm | 2 | 2 | 0 | 18.18 |
| Optimization (Linear programming, Mixed Integer programming) | 3 | 0 | 3 | 27.27 |
| Simulation | 4 | 2 | 2 | 36.36 |
| Multi-criteria decision making and Analytical hierarchy Process | 2 | 0 | 2 | 18.18 |
| Total | 11 | 4 | 7 | |

Table 4b. Classification based on OR methods

The classification based on OR methods clearly show that very few methods are used in the current area of research. The use of simulation and optimization is higher compared to other OR methods used. The use of methods like genetic algorithms is reported to have problems of computational complexity and local maxima how ever better versions of genetic algorithms relevant to the problem may give near optimal solutions in many cases. In the context of selection among multiple choices based on multiple factors/criteria multi – criteria decision making methods are very useful but almost all multi- criteria methods suffer from the problem of deciding the correct weights for the factors/ criteria under consideration. The use of methods like analytical hierarchy process is enormous when it is needed to rank the service providers from the available pool based on the user preferred criteria. The case of web service supply chains considering all the participating entities and optimizing for all to reach their objectives OR methods like goal programming, dynamic programming and other methods might be of great help in taking the problem towards solution.

E. Classification based on attributes

The need for non – functional attributes or QoS in a web environment is enormous due to multiple competitors providing the same functionality. The QoS attributes act as order winners for the service providers. There are a large number of QoS attributes and the QoS attribute most important for a service is dependent on the user requirements and the type of the service itself. The QoS attributes are how ever considered individually in the literature rather than being considered together. The classification based on the non – functional attributes is presented below in table 5a as shown. The classification itself shows the various QoS attributes possible for web services.

| Attribute | Total | Conference | Journal | % over total |
| --- | --- | --- | --- | --- |

|  |  | Papers | Papers | number of papers |
|---|---|---|---|---|
| Reliability | 11 | 8 | 3 | 18.97 |
| Latency | 2 | 1 | 1 | 3.45 |
| Availability | 7 | 4 | 3 | 12.07 |
| Execution time | 1 | 1 | 0 | 1.72 |
| Security | 2 | 1 | 1 | 3.45 |
| Safety | 1 | 1 | 0 | 1.72 |
| Throughput time | 3 | 2 | 1 | 5.17 |
| Response time | 2 | 1 | 1 | 3.45 |
| Time | 8 | 4 | 4 | 13.79 |
| Performance | 3 | 3 | 0 | 5.17 |
| Usability | 2 | 2 | 0 | 3.45 |
| Reputation | 2 | 2 | 0 | 3.45 |
| Responsiveness | 1 | 0 | 1 | 1.72 |
| Completion time | 1 | 1 | 0 | 1.72 |
| Mean peak period latency | 1 | 1 | 0 | 1.72 |
| Success of completion | 3 | 1 | 2 | 5.17 |
| Cost | 4 | 1 | 3 | 6.90 |
| Accessibility | 1 | 1 | 0 | 1.72 |
| Capacity | 1 | 1 | 0 | 1.72 |
| MTBF (Mean time before failure), MTTR (Mean time to recover) | 2 | 2 | 0 | 3.45 |
| Total | 58 | 38 | 20 |  |

Table 5a. Classification based on QoS attributes

The table clearly points out that there has not been even consideration of the QoS attributes as far as literature is concerned. The major research literature relating to QoS attributes focus on few parameters like reliability, availability and time in general. The literature how ever has each of these terms defined and represented mathematically.

| Attributes | Author(s) |
|---|---|
| Accessibility | [17], [18] |
| Availability | [17],[18], [19], [20], [21], [22], [23], [24], [25], [26], [39] |
| Capacity | [17] |
| Completion time | [27] |
| Cost | [21], [25], [33] |
| Execution time | [25] , [39] |
| Latency | [28] |
| Mean peak period latency | [27] |
| Performance | [17], [23], [26] [29] |
| Realiabilty | [17], [19], [20], [25], [27], [30], [31], [32] |
| Reputation | [29] ,[33],[39] |
| Response time | [19], [20], [24] |
| Security/Safety | [9], [20], [19] ,[32] ,[33] |
| Success of completion | [18], [27] |
| Throughput time | [19], [20] |
| Time | [21] ,[23], [33] , [34],[36], [37] |
| Usability | [17] [35] |
| Work load | [19] |

Table 5b QoS attributes literature

The table 5b points out the need to consider the QoS attributes more rigorously rather than concentrating on selected few QoS attributes. The need to concentrate evenly on all

QoS attributes is because these QoS attributes can be grouped into customer related and service provider related QoS attributes as shown below in table 5c.

| Customer | Service provider |
| --- | --- |
| Reliability | Success of completion Mean time before failure/ MTBF |
| Availability | Security |
| Usability | Safety |
| Accessibility | Capacity |
| Reputation | Performance |
|  | Cost |
| Time | Throughput time |
|  | Response time/Latency |
|  | Completion time |
|  | Execution time |

Table 5c. Grouping of QoS attributes into client and service provider related QoS

The QoS attributes on client column shows the QoS attributes that the clients considered being most important to select a service and the QoS attributes in the service provider side show how the client perceived QoS translates to the service provider. The research making use of this relationship does not exist to the best of our knowledge. This is a huge research gap that might answer many QoS related questions in the web service supply chains context.

IV. Research gaps

There is a huge gap in terms of the concept of web services supply chains, the QoS attributes, the mutual optimization based on the requirements of all participating entities. The gap needs to be addressed effectively and efficiently.

a. Conceptual gap:
There is a lack of definition of web services supply chain concept though there are definitions for many of the enablers and drivers of the web services supply chains. The defining of roles in a web services supply chains, the interaction among the entities and the issues related to this kind of pure service supply chain.

b. QoS / non – functional requirements gap:

The need for quality in web services supply chains is essential in order to differentiate from the crowd as pointed out earlier. The entities participating in a particular service delivery is not known so considering QoS attributes individually might night help since the web service supply chain is affected by individual bad performance to a higher degree than that of a static/ semi – static supply chains. A way of consolidating the QoS attributes is necessary for fixing minimum QoS requirements that each participating entity must satisfy and for easier QoS calculations.

c. Mutual optimization methods:

The mutual optimization for all entities considering each one's constraints, requirements, capabilities and considering how each entity's good / bad performance adds/ affects the others is needed. This kind of optimization is very complicated since it involves multiple view points and requirements. Some suitable methods for this kind of optimization is to be put forward to resolve this gap.

V. Conclusion

In this paper we presented a through literature review considering the area of web services supply chains and the need for QoS optimization in such supply chains. The gaps in various dimensions such as conceptual gap, QoS gap and the method gap are identified and pointed out. The current methods used, the QoS attributes considered and various other dimensions of the literature are classified and presented clearly for understanding the need for considering this new area of research. The defining of the characteristics, consolidation of QoS attributes for the web services supply chains and the mutual optimization of the same will be our future work.